\documentclass[%
 reprint,
 amsmath,amssymb,
 aps,
 pra,
]{revtex4-1}

\usepackage{graphicx}
\usepackage{dcolumn}
\usepackage{bm}

\begin{document}

\preprint{APS/PRA}

\title{Diffraction of Bessel beams on 2D amplitude gratings -- a new branch in the Talbot effect study}

\author{I.A. Kotelnikov}
\affiliation{Budker Institute of Nuclear Physics SB RAS, 630090 Novosibirsk, Russia}
\affiliation{Novosibirsk State University, 630090 Novosibirsk, Russia}
\author{O.E. Kameshkov}
\affiliation{Budker Institute of Nuclear Physics SB RAS, 630090 Novosibirsk, Russia}
\affiliation{Novosibirsk State University, 630090 Novosibirsk, Russia}
\author{B.A. Knyazev}
\email[e-mail address: ]{B.A.Knyazev@inp.nsk.su}
\affiliation{Budker Institute of Nuclear Physics SB RAS, 630090 Novosibirsk, Russia}
\affiliation{Novosibirsk State University, 630090 Novosibirsk, Russia}

\renewcommand{\vec}[1]{\mathbf{#1}}

\date{\today}

\begin{abstract}
In this paper, an analytical theory for the diffraction of a Bessel beam of arbitrary order $J_l(\kappa r)$ on a 2D amplitude grating is presented. The diffraction pattern in the main and fractional Talbot planes under certain conditions is a lattice of annular microbeams, the diameters of which depend on the grating period, the illuminating beam diameter, the number of the Talbot plane, and the topological charge $l$. For the rings near the optical axis, the latter reproduces $l$ of the illuminating beam. Experiments carried out on the Novosibirsk free electron laser at a wavelength $\lambda = 141$~$\mu$m using gratings with hole diameters of down to $d \approx 2\lambda $, as well as, the numerical simulations, well support the theory. Since the Laguerre-Gaussian beams can be represented as a superposition of Bessel beams, results of this paper can be applied to the analysis of the Talbot effect with the Laguerre-Gaussian beams.
\end{abstract}


\maketitle

\section{Introduction}
\label{sec:Introduction}

In 1836, H.~F.~Talbot discovered \cite{Talbot1836}, that behind a periodic linear grating, illuminated by a distant emitting point of intense sunlight, a system of colored lines with a period equal to the grating period was observed in planes at equal distances. Since the light source was non-monochromatic, he observed color lines and stripes alternating with increasing distance. The patterns described by him in the article can now be easily explained, known that the distance between the planes of self-imaging (the Talbot length) depends on the wavelength. In the same paper, Talbot investigated the diffraction of light on a two-dimensional periodic lattice of small holes in copper foil, and also discovered the phenomenon of self-imaging, although he did not describe the latter in sufficient detail. J.~W.~Strutt (Lord Rayleigh) studied in detail the diffraction of monochromatic light on absorbing gratings in 1881 \cite{Rayleigh1881}. He measured the distances between the planes of self-imaging, and gave a mathematical description of this phenomenon as light diffraction in the near field. He derived analytically the distances between the planes of exact self-imaging, which is the Talbot length equal to $ z_T= 2p^2/\lambda $, where $p$ is the grating period and $\lambda$ is the wavelength. It is worth noting that in the experiments he also observed images of the grating in the planes at $z_T/2$ \cite{Rayleigh_length}.

After about half a century, interest in this phenomenon was revived, which coincided with the invention of lasers \cite{Maiman1960} and the possibility of obtaining coherent monochromatic radiation beams, which gave an additional impetus to further study of the effect. Among many theoretical papers, references to which can be found in reviews \cite{Patorsky1989,Cowley1995,Wen2013}, mention should be made of papers \cite{Cowley-1,Cowley-2,Winthrop1965,Montgomery1967}, in which the modern theory of the Talbot effect was formulated. In papers  \cite{Cowley-1,Cowley-2,Winthrop1965} it was shown that between the planes of the exact self-image of the grating, called by the authors the Fourier images, there are planes in which the grating images with a higher spatial frequency (Fresnel images) are observed. In the modern terminology, these planes are called the main and fractional Talbot planes, in accordance with the formula that describes their position,
\begin{equation}
\label{FractTalbotPlanes}
\begin{array}{*{20}{c}}
  {z(j,u,v) = {z_T}\left( {j + \beta } \right),}&{\beta  = \frac{u}{v},}&{{z_T} = \frac{{2{p^2}}}{\lambda },}
\end{array}
\end{equation}
where $j=0,1,2,3...$  and $u$ are whole numbers,  $v=1,2,3...$ is a natural  number,  $u<v$, and $z_T$ is the Talbot distance. The planes with $u=0$ and $j > 0$ are the main Talbot planes.  With $u/v=1/2$ (``half-Talbot planes''), the image of the grating is also exactly reproduced but shifted along both coordinates by half the period. In planes with $u/v=1/4$ and $3/4$ (``quarter-Talbot planes''), there are observed grating images of double spatial frequency, whereas at $u/v=1/6,1/3,2/3$, and $5/6$, the patterns have a triple spatial frequency.


Furthers studies have also shown that the appearance of self-image does not require the full periodicity of the object, but rather its quasi-periodicity, and the Talbot effect itself is a special case of the more general Montgomery effect \cite{Montgomery1967}. Lohmann in \cite{Lohmann2005} systematized the previous studies of the fundamental and fractional Talbot effects and the Montgomery effect. He described the ``Montgomery fractional effect'' and gave an elegant explanation of the relationship between these effects using graphical analysis of the effect in the spectrum of spatial frequencies. Since the content of our work will be connected with the study of diffraction on periodic gratings, we will not dwell on this question.

The overwhelming majority of studies on the Talbot effect were performed with gratings  illuminated with the plane and spherical waves. Only in recent years, several publications have appeared in which the diffraction of beams with orbital angular momentum on linear and two-dimensional gratings has been studied. In the paper \cite{Panthong16},  the Talbot effect on a linear grating was applied to the detection of topological charge of a vortex Laguerre-Gaussian beam. In our paper \cite{Knyazev2018OE}, a first experimental and numerical study of the diffraction of vortex Bessel beams on two-dimensional gratings of circular openings was carried out. It was discovered that a regular lattice of annular vortex microbeams with a period equal to the grating period appeared in the Talbot planes. Details of the numerical calculations can be found in \cite{IzvRAN}. An analytical expression for images in the main Talbot planes, when a 2D grating was illuminated by a vortex Bessel beam, was given in brief communication \cite{TERA}. Analytical expressions for images in the main Talbot planes for phase and amplitude two-dimensional gratings illuminated by vortex Gaussian beams were published in \cite{RASOULI}.

In \cite{Knyazev2018OE} we attributed the image of the grating holes in the form of rings to the illumination of the grating by a vortex beam. Subsequent analysis, however, led us to the conclusion that the reason for the appearance of rings is not the twist of the incident beam, but its mode composition, namely that the Bessel beams were a superposition of conically converging plane waves. Indeed, we observed the rings experimentally when the grating was illuminated by a zero-order Bessel beam. The effect of twisting is that in our case these rings have the same topological charge as the incident beam. According to above-mentioned work \cite{RASOULI}, in the case of a vortex Gaussian beam, other regularities are observed for the topological charge of the rings in the main Talbot planes. In this paper, we present an analytical study of the Talbot effect when a 2D grating is illuminated by a Bessel beam of arbitrary order and derive the expressions for the main and fractional Talbot planes. The obtained expressions are confirmed by the results of experiments and numerical calculations.

\section{Transmission of a plane wave through a two-dimensional grating}
\label{sec:plane_wave_transmission}

Let a plane wave be incident normally on a two-dimensional periodic grating. Directly behind the grating, in the $ z = 0 $ plane, the wave field can be represented as an infinite series
\begin{equation}
\label{2:01}
E(x,y,0 + ) = \sum\limits_{mn} {{E_{mn}}} \exp \left[ {i\frac{{2\pi }}{p}\left( {m{\mkern 1mu} x + n{\mkern 1mu} y} \right)} \right],
\end{equation}
where $ p $ is the grating period, and the coefficients $ E_{mn} $ depend on the shape of the holes in the grating. We calculate the Fourier amplitude of this field using the well-known representation of the delta function:
\begin{equation}
\label{2:02}
\int_{ - \infty }^\infty  {{e^{ - ikx}}} dx = 2\pi {\mkern 1mu} \delta (k).
\end{equation}
As a result, we find
\begin{multline}
\label{2:03}
  E({k_x},{k_y}) = \iint {dx}dy{\mkern 1mu} E(x,y,0 + ){e^{ - i{k_x}x - i{k_y}y}}= \\
  = {(2\pi )^2}\sum\limits_{mn} {{E_{mn}}} {\mkern 1mu} \delta ({k_x} - 2\pi m/p){\mkern 1mu} \delta ({k_y} - 2\pi n/p).
\end{multline}
The field in the space behind the grating is
\begin{equation}
\label{2:04}
E(x,y,z) = \iint {\frac{{d{k_x}{\text{d}}{k_y}}}{{{{(2\pi )}^2}}}}{\mkern 1mu} E({k_x},{k_y}){\mkern 1mu} {{\text{e}}^{i{k_x}x + i{k_y}y + i{k_z}z}},
\end{equation}
where
\begin{equation}
\label{2:05}
{k_z} = \sqrt {{k^2} - k_x^2 - k_y^2}  \approx k - \frac{{k_x^2}}{{2k}} - \frac{{k_y^2}}{{2k}}.
\end{equation}
Substituting \eqref{2:03} into \eqref{2:04} and using the definition of the Talbot length in  \eqref{FractTalbotPlanes}, we get
\begin{multline}
\label{2:06}
E(x,y,z) = {e^{ikz}}\sum\limits_{mn} {{E_{mn}}} \exp \left[ {i\frac{{2\pi }}{p}\left( {mx + ny} \right)} \right]  \times
 \\  \times \exp \left[ { - i\frac{{2\pi }}{{{z_T}}}\left( {{m^2} + {n^2}} \right)z} \right].
\end{multline}
Comparing \eqref {2:06} and \eqref {2:01}, it is easy to see that in the main  Talbot planes
\begin{equation}
\label{2:07}
{z_j} = {z_T}j = \frac{{2{p^2}}}{\lambda }{\mkern 1mu} j\qquad j = 1,2,3, \ldots
\end{equation}
images of the grating are repeated:
\begin{equation}
\label{2:08}
E(x,y,{z_j}) = {e^{ik{z_j}}}\sum\limits_{mn} {{E_{mn}}} \exp \left[ {i\frac{{2\pi }}{p}\left( {mx + ny} \right)} \right].
\end{equation}
Recall that the phase factor ${e^{ik{z_j}}}$ does not affect the wave intensity. The same result can be obtained by solving the Kirchhoff integral.

\section{Grating with round holes}
\label{Grating_with_round_holes}

Let us find the Fourier series expansion for the transmission function of a two-dimensional grating with a period of $p$ and round holes with a diameter of $d = 2R<p$. For such a grating, the transmission function is written as
\begin{equation}
\label{3:01}
h(\vec x) = \sum\limits_{mn} \theta  \left( {R - |\vec x - {{\vec p}_{mn}}|} \right),
\end{equation}
where for the two-dimensional vectors the notations
\begin{equation}
\label{3:02}
\begin{array}{*{20}{c}}

  {\vec x = \{ x,y\} }&{{{\vec p}_{mn}} = \{ mp,np\} ,}
\end{array}
\end{equation}
are introduced, and the Heaviside function  $\theta(\xi)$ is equal to $0$ at $\xi<0$ and $1$ at $\xi>0$.
Let us calculate the Fourier integral
\begin{equation}
\label{3:03}
h({{\vec k}_ \bot }) = \iint {{d^2}x}{e^{ - i{{\vec k}_ \bot } \cdot \vec x}}h(\vec x),
\end{equation}
where $\vec{k}_{\bot}=\{k_{x},k_{y}\}$ denotes a two-dimensional vector. Taking into account the periodicity of the function $h(\vec{x})$ and replacing $\vec{x}$ with $\vec{x}'+\vec{p}_{mn}$,  we convert the integral to the sum of the series
\begin{equation}
\label{3:04}
h({{\vec k}_ \bot }) = \sum\limits_{mn} {{e^{ - i{{\vec k}_ \bot } \cdot {{\vec p}_{mn}}}}} \iint\limits_{|x'| < R} {{d^2}x'}{e^{ - i{{\vec k}_ \bot } \cdot \vec x'}}.
\end{equation}
It is important that the internal integral here does not depend on the summation indices and can be expressed through the Bessel function $ {{\text{J}}_1} $, therefore
\begin{equation}
\label{hkperp}
h({{\vec k}_ \bot }) = \frac{{2\pi R}}{{{k_ \bot }}}{\mkern 1mu} {J_1}({k_ \bot }R)\sum\limits_{mn} {{e^{ - i{{\vec k}_ \bot } \cdot {{\vec p}_{mn}}}}}
\end{equation}
For a rectangular grating, the sum $ \sum_{mn} $ is expressed in terms of functions well-known in the diffraction theory:
\begin{multline*}
\label{3:05}
\sum\limits_{m =  - M}^M {\sum\limits_{n =  - N}^N {{e^{ - i\alpha m - i\beta n}}} }=
\\
 = \frac{{\sin \left[ {\left( {M + 1/2} \right)\alpha } \right]}}{{\sin [\alpha /2]}}\frac{{\sin \left[ {\left( {N + 1/2} \right)\beta } \right]}}{{\sin [\beta /2]}},
\end{multline*}
where $\alpha = k_{x}p$ and  $\beta = k_{y}p$.   It is noticeably nonzero only near the points $\vec{k}_{\bot}=\vec{q}_{mn}$, where (see.~Eq.~\ref{2:01})
\begin{equation}
\label{q_mn} \vec{q}_{mn}=\{2\pi m/p,2\pi n/p\}.
\end{equation}
For an infinite grating, the sum of the series is calculated using the representation of the periodic delta function:
\begin{equation}
\label{3:06}
\sum\limits_{m =  - \infty }^\infty  {\left( {2\pi } \right)} \delta (x - 2\pi m) = \sum\limits_{n =  - \infty }^\infty  {{e^{inx}}} .
\end{equation}
With its help, from \eqref{hkperp} we get
\begin{multline}
\label{3:07}
   h({{\vec k}_ \bot }) = \frac{{2\pi R}}{{{k_ \bot }}}{\mkern 1mu} {J_1}({k_ \bot }R)\sum\limits_{mn} {{{\left( {2\pi /p} \right)}^2}} \delta ({{\vec k}_ \bot } - {{\vec q}_{mn}})=
    \\=
    \frac{{2\pi {R^2}}}{{{p^2}}}\sum\limits_{mn} {\frac{{{J_1}({q_{mn}}R)}}{{{q_{mn}}R}}} {\mkern 1mu} {\left( {2\pi } \right)^2}\delta ({{\vec k}_ \bot } - {{\vec q}_{mn}}).
\end{multline}
Performing the inverse Fourier transform, we find the representation of the transmission function as a Fourier series:
\begin{multline}
\label{3:08}
h(\vec x) = \iint {\frac{{{d^2}{k_ \bot }}}{{(2\pi )}}}{\mkern 1mu} h({{\vec k}_ \bot }){\mkern 1mu} {e^{i{{\vec k}_ \bot } \cdot \vec x}} =
\\
=\frac{{2\pi {R^2}}}{{{p^2}}}\sum\limits_{mn} {\frac{{{J_1}({q_{mn}}R)}}{{{q_{mn}}R}}} {e^{i{{\vec q}_{mn}} \cdot \vec x}}.
\end{multline}

\section{Diffraction of the Bessel wave}
\label{Diffraction_of_the_Bessel_wave}
Let us try to expand the theory of the Talbot effect for the case when a cylindrical Bessel wave illuminates a two-dimensional periodic grating. The cylindrical Bessel wave \begin{equation}
    \label{4:01}
{E_{l\kappa {k_z}}} = {J_l}(\kappa r){e^{il{\phi _r} + i{k_z}z}}
\end{equation}
in the cylindrical coordinate system $\{r,\phi_{r},z\}$ is characterized by the wave numbers $\kappa$, $l$, $k_{z}=\sqrt{k^{2}-\kappa^{2}}$ (see, e.~g. \cite{Serbo18}). The wave number $l$ is often called ``the topological charge.'' The Bessel wave can be represented as a superposition of plane waves.
\begin{equation}
\label{4:02}
{E_{l\kappa {k_z}}} = \iint {\frac{{{d^2}{k_ \bot }}}{{{{(2\pi )}^2}}}}{\mkern 1mu} {a_{\kappa l}}({{\vec k}_ \bot }){e^{i\vec k \cdot \vec r}},
\end{equation}
where $\vec{k}_{\bot}=\{k_{\bot}\cos\phi_{k}, k_{\bot}\sin\phi_{k}\}$, $\vec{k}=\{\vec{k}_{\bot},k_{z}\}$ and the Fourier amplitude is written as
\begin{equation}
\label{4:03}
{a_{\kappa l}}({{\vec k}_ \bot }) = {i^{ - l}}{e^{il{\phi _k}}}{\mkern 1mu} {\mkern 1mu} \frac{{2\pi }}{\kappa }{\mkern 1mu} \delta \left( {{k_ \bot } - \kappa } \right).
\end{equation}
The plane waves in such a superposition lie on a cone with an vertex angle
\begin{equation}
\label{4:04}
{\theta _k} = \arctan (\kappa /{k_z}).
\end{equation}
By calculating the Fourier amplitude of the field, the incident cylindrical Bessel wave generates directly behind the grating, we can apply the method used in section \ref{sec:plane_wave_transmission}  to find the diffracted field of the plane wave in the Talbot planes.

First, we will take into account that the fields directly in front of the lattice and behind it are related by the expression
\begin{equation}
\label{4:05}
{E_ + }(\vec x) = h(\vec x){\mkern 1mu} {E_ - }(\vec x),
\end{equation}
where $h(\vec{x})$ is the grating transmission coefficient. Then we will express the Fourier amplitude of the field $E_{+}(\vec{x})$ in terms of the Fourier amplitudes of the transmission function of the grating $h(\vec{x})$ and the incident cylindrical wave $E_{-}(\vec{x})$:
\begin{multline*}
{E_ + }({{\bf{k}}_ \bot }) = \iint{d^2}x{\kern 1pt} {e^{ - i{{\bf{k}}_ \bot }\cdot{\bf{x}}}}h({\bf{x}}){E_ - }({\bf{x}}) = \\
\quad \quad \quad  = \iint{d^2}x{\kern 1pt} {e^{ - i{{\bf{k}}_ \bot }\cdot{\bf{x}}}}\iint\frac{{{d^2}{{k'}_ \bot }}}{{{{(2\pi )}^2}}}{e^{i{{{\bf{k'}}}_ \bot }\cdot{\bf{x}}}}h({{{\bf{k'}}}_ \bot }) \times \\
\quad \quad \quad \quad \quad \quad \quad \quad \quad   \times \iint\frac{{{d^2}{{k''}_ \bot }}}{{{{(2\pi )}^2}}}{e^{i{{{\bf{k''}}}_ \bot }\cdot{\bf{x}}}}{E_ - }({{{\bf{k''}}}_ \bot } ).
\end{multline*}
Since
\begin{equation*}
\label{4:07}
\iint {{d^2}}x{e^{ - i({{\vec k}_ \bot } - {{\vec k'}_ \bot } - {{\vec k''}_ \bot }) \cdot \vec x}} = {(2\pi )^2}\delta ({{\vec k}_ \bot } - {{\vec k'}_ \bot } - {{\vec k''}_ \bot }),
\end{equation*}
we conclude that
\begin{equation}
\label{4:08}
{E_ + }({{\vec k}_ \bot }) = \iint {\frac{{{d^2}{{\vec k'}_ \bot }}}{{{{(2\pi )}^2}}}}h({{\vec k'}_ \bot }){\mkern 1mu} {E_ - }({{\vec k}_ \bot } - {{\vec k'}_ \bot }).
\end{equation}

The integral of \eqref{4:08} can be calculated analytically by substitution of the Fourier amplitude $h(\vec{k}_{\bot}')$ from \eqref{3:07}:
\begin{equation}
\label{4:09}
{E_ + }({{\vec k}_ \bot }) = \frac{{2\pi {R^2}}}{{{p^2}}}\sum\limits_{mn} {\frac{{{J_1}({q_{mn}}R)}}{{{q_{mn}}R}}} {E_ - }({{\vec k}_ \bot } - {{\vec q}_{mn}}).
\end{equation}
In our case, the value $ E_{-} (\vec{k}_{\bot}) $ should be replaced with the function $ a(\vec{k}_{\bot}) $ from \eqref{4:03}. As a result, we obtain the following expression for the Fourier amplitude of plane waves in a diffracted field
\begin{multline}
\label{4:10}
{E_ + }({{\vec k}_ \bot }) = \frac{{2\pi {R^2}}}{{{p^2}}}\sum\limits_{mn} {\frac{{{J_1}({q_{mn}}R)}}{{{q_{mn}}R}}} {\mkern 1mu}  \times
    \\
\times {i^{ - l}}{e^{il{{\phi '}_k}}}\frac{{2\pi }}{\kappa }{\mkern 1mu} \delta (|{{\vec k}_ \bot } - {{\vec q}_{mn}}| - \kappa ),
\end{multline}
where it is necessary to distinguish the indices $ m $ and $ n $, over which the summation is performed, and the azimuth number $ l $ of the incident wave. We also indicate that $ \phi_{k}'$ means the azimuthal angle of the two-dimensional vector $ \vec{k}_{\bot}' = \vec{k}_{\bot} - \vec{q}_{mn} $. Due to the presence of the delta function in the expression \eqref {4:10}, only vectors $ \vec{k}_{\bot} $ such that the length of the vector $ \vec{k}_{\bot}'$ is $ \kappa$, i.e.
\begin{equation*}
\label{4:11}
{{\vec k}_ \bot } - {{\vec q}_{mn}} = \{ \kappa \cos {{\phi '}_k},\kappa \sin {{\phi '}_k}\},
\end{equation*}
contribute to the field behind the grating. Each member of the series of \eqref{4:10} with given indices $ m $ and $ n $ is a superposition of plane waves lying on a cone with a vertex angle $ \theta_{k} = \arctan(\kappa/k_{z} ) $ around the direction of the vector $ \{\vec{q}_{mn},k_{z}\} $ to the center of the image of the corresponding hole. This angle is exactly equal to the angle at the vertex of the cone of \eqref{4:04} in the decomposition into plane waves of the original Bessel wave. As in the incident Bessel wave, these plane waves are uniformly distributed on such a cone, since the terms of the series depend on the angle $ \phi_{k}'$ only through the phase factor $ {e^{il{\phi _{k'}}}}$.

The resulring expression for the spectrum of spatial frequencies indicates that when the grating is illuminated with a Bessel beam, it can be expected that, unlike the classical Talbot effect, the ``images'' of the holes will turn into rings. This effect was originally discovered in experiment and numerical simulation \cite{Knyazev2018OE}.
For a complete proof of this statement, one would have to calculate the integral of \eqref{2:04} for the field of \eqref{4:10}:
\begin{equation}
\label{4:12}
E(\vec x,z) = \iint {\frac{{{d^2}{k_ \bot }}}{{{{(2\pi )}^2}}}}{\mkern 1mu} {E_ + }({{\vec k}_ \bot }){\mkern 1mu} {e^{i{{\vec k}_ \bot } \cdot \vec x + i{k_z}z}},
\end{equation}
where $k_{z}\approx k-k_{\bot}^{2}/2k$. Substituting here the expression \eqref{4:10} for $ E_{+}(\vec{k}_{\bot}) $ and performing the replacement of the integration variable $ \vec{k}_{\bot}=\vec{k}_{\bot}'+\vec{q}_{mn}$, we get
\begin{multline}
    \label{4:13}
E(\vec x,z) = \frac{{2\pi {R^2}}}{{{p^2}}}\sum\limits_{mn} {\frac{{{J_1}({q_{mn}}R)}}{{{q_{mn}}R}}} {\mkern 1mu} {e^{i{{\vec q}_{mn}} \cdot \vec x + ikz}} \times
    \\   \times \iint {\frac{{d{{k'}_ \bot }}}{{{{(2\pi )}^2}}}}{\mkern 1mu} {i^{ - l}}{e^{il{{\phi '}_k}}}\frac{{2\pi }}{\kappa }{\mkern 1mu} \delta ({{k'}_ \bot } - \kappa ){e^{i{{\vec k'}_ \bot } \cdot \vec x}} \times
    \\  \times {e^{ - i|{{\vec k'}_ \bot } + {{\vec q}_{mn}}{|^2}z/2k}}.
\end{multline}
Expression~(\ref{4:13}) is a general solution for the electric field behind a periodic grating with round holes of arbitrary radii illuminated with a Bessel beam of the $l$th order. If it were not for the factor ${e^{ - i|{{\vec k}_{ \bot '}} + {{\vec q}_{mn}}{|^2}z/2k}}$ in the last line, in the internal integral one could recognize the superposition of plane waves, which makes up a cylindrical Bessel wave.

\section{Diffraction patterns in the Talbot planes}

Let us, first, concentrate on the calculation of this integral in the main and fractional Talbot planes $z=(j+\beta)z_{T}$. Since the contribution to the integral is made only by such vectors $\vec{k}_{\bot}'$, for which $|\vec{k}_{\bot}'|=\kappa $, and ${\left| {{q_{mn}}} \right|^2}/2k = 2\pi  \cdot \left( {{m^2} + {n^2}} \right)/{z_T}$, then for the last factor in (\ref{4:13}) we obtain:
\begin{multline}\label{5:01}
  {e^{ - i|{{\vec k'}_ \bot } + {{\vec q}_{mn}}{|^2}z/2k}} ={e^{ - i{\kappa ^2}(j + \beta ){z_T}/2k}} \times
   \\
    \times {e^{ - i{\kern 1pt} ({{\vec k'}_ \bot } \cdot {{\vec q}_{mn}}){\kern 1pt} (j + \beta ){z_T}/k}}{e^{ - 2\pi i\left( {{m^2} + {n^2}} \right)\left( {j + \beta } \right)}}.
\end{multline}
The factor ${e^{ - i{\kappa ^2}(j + \beta ){z_T}/2k}}{e^{ - 2\pi i\left( {{m^2} + {n^2}} \right)\left( {j + \beta } \right)}}$ can be taken out of the integral, since it does not depend on $\vec{k}_{\bot}'$. Therefore, the desired integral is transformed to
\begin{multline}\label{5:02}
  E(\vec x,j,\beta ) = \frac{{2\pi {R^2}}}{{{p^2}}}\sum\limits_{mn} {\frac{{{J_1}({q_{mn}}R)}}{{{q_{mn}}R}}} {\mkern 1mu} {e^{i{{\vec q}_{mn}} \cdot \vec x + ikz}} \times
   \\
   \times {e^{ - i{\kappa ^2}(j + \beta ){z_T}/2k}}{e^{ - 2\pi i\left( {{m^2} + {n^2}} \right)\left( {j + \beta } \right)}} \times
   \\
 \times \iint {\frac{{{d^2}{{k'}_ \bot }}}{{{{(2\pi )}^2}}}}{\mkern 1mu} {i^{ - l}}{e^{il{{\phi '}_k}}}\frac{{2\pi }}{\kappa }{\mkern 1mu} \delta ({{k'}_ \bot } - \kappa )\times
 \\
 \times{e^{i{{\vec k'}_ \bot } \cdot \vec x}}{e^{ - i{\kern 1pt} ({{\vec k'}_ \bot } \cdot {{\vec q}_{mn}}){\kern 1pt} (j + \beta ){z_T}/k}}.
\end{multline}
In the limit of the grating with a small radius of holes, we obtain approximately
\begin{equation*}
   \frac{{{J_1}\left( {{q_{mn}}R} \right)}}{{{q_{mn}}R}} \approx \frac{1}{2}.
\end{equation*}
Then applying summation by the Poisson rule to \eqref{5:02} (see Appendix~\ref{app:Poisson}) we get
\begin{multline}\label{5:03}
  E(\vec x,j,\beta ) = \frac{{\pi {R^2}}}{{{p^2}}}{e^{ - i{\kappa ^2}(j + \beta ){z_T}/2k + ikz}} \times
   \\
  \times \sum\limits_{m'n'} {\iint {dmdn}{e^{ - 2\pi im'm - 2\pi in'n}}{e^{i{{\vec q}_{mn}} \cdot \vec x}}{e^{ - 2\pi i\left( {{m^2} + {n^2}} \right)\left( {j + \beta } \right)}}}  \times
   \\
   \times \iint {\frac{{{d^2}{{k'}_ \bot }}}{{{{(2\pi )}^2}}}}{\mkern 1mu} {i^{ - l}}{e^{il{{\phi '}_k}}}\frac{{2\pi }}{\kappa }{\mkern 1mu} \delta ({{k'}_ \bot } - \kappa )\times
   \\
   \times {e^{i{{\vec k'}_ \bot } \cdot \vec x}}{e^{ - i{\kern 1pt} ({{\vec k'}_ \bot } \cdot {{\vec q}_{mn}}){\kern 1pt} (j + \beta ){z_T}/k}}{\mkern 1mu}.
\end{multline}

In the resulting expression, we move from the integration over $m$ and $n$ to the integration over ${d^2}{q_{mn}}$. Taking into account that ${dmdn = {p^2}{d^2}{q_{mn}}/{{\left( {2\pi } \right)}^2}}$ and ${{\vec q}_{mn}} \cdot {{\vec p}_{m'n'}} = 2\pi im'm + 2\pi in'n$, we get the general solution of the problem
\begin{multline}\label{5:04}
  E(\vec x,j,\beta ) = \pi {R^2}{e^{ - i{\kappa ^2}(j + \beta ){z_T}/2k + ikz}} \times
  \\
   \times \sum\limits_{m'n'} {\iint {\frac{{{d^2}{{k'}_ \bot }}}{{{{(2\pi )}^2}}}}{\mkern 1mu} {i^{ - l}}{e^{il{{\phi '}_k}}}\frac{{2\pi }}{\kappa }{\mkern 1mu} \delta ({{k'}_ \bot } - \kappa ){e^{i{{\vec k'}_ \bot } \cdot \vec x}}}  \times
   \\
   \times  \iint {\frac{{{d^2}{q_{mn}}}}{{{{(2\pi )}^2}}}}{e^{i{{\vec q}_{mn}}\left[ {\vec x - {{\vec p}_{m'n'}} - {{\vec k'}_ \bot }(j + \beta ){z_T}/k} \right]}}]\times \\
   \times {e^{ - 2\pi i\left( {{m^2} + {n^2}} \right)\left( {j + \beta } \right)}}.
\end{multline}
If there were no the last multiplier (${e^{ - 2\pi i\left( {{m^2} + {n^2}} \right)\left( {j + \beta } \right)}}$), the integral
\begin{multline}\label{5:05}
 I = \iint {\frac{{{d^2}{q_{mn}}}}{{{{(2\pi )}^2}}}}{e^{i{{\vec q}_{mn}}\left[ {\vec x - {{\vec p}_{m'n'}} - {{\vec k'}_ \bot }(j + \beta ){z_T}/k} \right]}} \times
 \\
 \times{e^{ - 2\pi i\left( {{m^2} + {n^2}} \right)\left( {j + \beta } \right)}}
\end{multline}
could be calculated easily. Thus, the solution has to be investigated for particular Talbot planes.
\subsection{Main Talbot planes}
We start from the  main Talbot planes ($\beta  = 0$):
\begin{equation}\label{5:06}
 \exp \left( { - 2\pi i\left( {{m^2} + {n^2}} \right)j} \right) = 1.
\end{equation}
The integral $I$ over ${{q_{mn}}}$ is expressed in terms of the delta function and we get
\begin{multline}\label{5:07}
    I = \iint {\frac{{{d^2}{q_{mn}}}}{{{{(2\pi )}^2}}}{e^{i{{\vec q}_{mn}}\left( {\vec x - {{\vec p}_{m'n'}} - {{\vec k'}_ \bot }{z_T}j/k} \right)}}} =
   \\
   \delta \left( {\vec x - {{\vec p}_{m'n'}} - {{\vec k'}_ \bot }{z_T}j/k} \right).
\end{multline}
Finally,
\begin{multline}\label{5:08}
   E(\vec x,j,0) = \pi {R^2}{e^{ - i{\kappa ^2}j{z_T}/2k + ikz}} \times
  \\
  \times \sum\limits_{m'n'} {\iint {\frac{{{d^2}{{k'}_ \bot }}}{{{{(2\pi )}^2}}}}{\mkern 1mu} {i^{ - l}}{e^{il{{\phi '}_k}}}\frac{{2\pi }}{\kappa }{\mkern 1mu} \delta ({{k'}_ \bot } - \kappa ){e^{i{{\vec k'}_ \bot } \cdot \vec x}}}  \times
  \\
  \times \delta \left( {\vec x - {{\vec p}_{m'n'}} - {{\vec k'}_ \bot }{z_T}j/k} \right).
\end{multline}
It is seen that the pattern is a lattice of annular beamlets with centers at the points
\begin{equation}\label{5:09}
 \vec x = {{\vec p}_{m'n'}},
\end{equation}
corresponding to the coordinates of the grating openings,  with radii
\begin{equation}\label{5:10}
    {\rho_j} = \kappa j{z_T}/k = \kappa \frac{{2{p^2}}}{{\lambda k}}{\mkern 1mu} j = \kappa \frac{{{p^2}}}{\pi }{\mkern 1mu} j,
\end{equation}
and the same  topological charge as that of the illuminating beam.

\subsection{Half Talbot planes}
Now let us substitute into the general solution \eqref{5:04} the coefficient $\beta  = 1/2$  corresponding to the Talbot half-planes ($z=z_T(j+1/2)$). In this case
\begin{multline}\label{5:11}
    \exp \left[ { - 2\pi i\left( {{m^2} + {n^2}} \right)\left( {j + 1/2} \right)} \right]=
    \\ = \exp \left[ { - \pi i\left( {{m^2} + {n^2}} \right)} \right] =
   \\
    = \exp \left[ { - \pi i\left( {m + n} \right)} \right] = \exp \left[ { - i{{\vec p}_{11}}{{\vec q}_{mn}}/2} \right].
\end{multline}
Performing transformations similar to the transformations in \eqref{5:07}
\begin{multline}\label{5:12}
    I = \iint {\frac{{{d^2}{q_{mn}}}}{{{{(2\pi )}^2}}}{e^{i{{\vec q}_{mn}}\left( {\vec x - {{\vec p}_{m'n'}} - {{\vec k'}_ \bot }(j + 1/2){z_T}/k} \right) - {{\vec p}_{11}}/2}}} =
   \\
    = \delta \left( {\vec x - {{\vec p}_{m'n'}} - {{\vec p}_{11}}/2 - {{\vec k'}_ \bot }(j + 1/2){z_T}/k} \right),
\end{multline}
we get an expression similar to expression~\eqref{5:08}
\begin{multline}\label{5:13}
    E(\vec x,j,1/2) = \pi {R^2}{e^{ - i{\kappa ^2}(j + 1/2){z_T}/2k + ikz}} \times
    \\
    \times \sum\limits_{m'n'} {\iint {\frac{{{d^2}{{k'}_ \bot }}}{{{{(2\pi )}^2}}}}{\mkern 1mu} {i^{ - l}}{e^{il{{\phi '}_k}}}\frac{{2\pi }}{\kappa }{\mkern 1mu} \delta ({{k'}_ \bot } - \kappa ){e^{i{{\vec k'}_ \bot } \cdot \vec x}}}  \times
    \\
    \times \delta \left( {\vec x - {{\vec p}_{m'n'}} - {{\vec p}_{11}}/2 - {{\vec k'}_ \bot }(j + 1/2){z_T}/k} \right).
\end{multline}
From this equation it follows that the rings are shifted by half a period with some new radii
\begin{equation}\label{5:14}
   \begin{array}{*{20}{c}}
  {\vec x = {{\vec p}_{m'n'}} + {{\vec p}_{11}}/2},&{{\rho_{j,1/2}} = \kappa \frac{{{p^2}}}{\pi }{\mkern 1mu} \left( {j + \frac{1}{2}} \right)}.
   \end{array}
\end{equation}

\subsection{Fractional Talbot planes}
Rings can also occur at the fractional Talbot distances $z = {z_T}(j + u/v)$. The $j + u/v$-dependent factor ${e^{ - 2\pi i\left( {{m^2} + {n^2}} \right)\left( {j + u/v} \right)}}$ in  \eqref{5:04} is periodic in $m$ and $n$ with the period $v$ (see \eqref{FractTalbotPlanes}), which enables expansion of this expression into the discrete Fourier series \cite{Case2009}
\begin{multline}\label{5:15}
e^{ - 2\pi i\left( {m^2  + n^2 } \right)\left( {j + u/v} \right)}  = e^{ - 2\pi i\frac{u}
{v}\left( {m^2  + n^2 } \right)}  = \\ =\sum\limits_{s = 0}^{v - 1} {\sum\limits_{t = 0}^{v - 1} {a_{st} e^{ - 2\pi ism/v} e^{ - 2\pi itn/v} } }= \\ = \sum\limits_{s = 0}^{v - 1} {\sum\limits_{t = 0}^{v - 1} {a_{st} e^{ - i\vec q_{mn} \cdot\vec p_{st} /v} } }
\end{multline}
where
\begin{equation}\label{5:16}
 {a_{st}} = \frac{1}{{{v^2}}}\sum\limits_{m' = 0}^{v - 1} {\sum\limits_{n' = 0}^{v - 1} {{e^{ - 2\pi iu\left( {{{m'}^2} + {{n'}^2}} \right)/v}}} {e^{i{{\vec q}_{m'n'}} \cdot {{\vec p}_{st}}/v}}}.
\end{equation}
As a result, we obtain the expression for the electric field,
\begin{multline}\label{5:17}
   E(\vec x,j,\beta ) = \pi {R^2}{e^{ - i{\kappa ^2}(j + \beta ){z_T}/2k + ikz}} \times
    \\
    \times \sum\limits_{m'n'} {\iint {\frac{{{d^2}{{k'}_ \bot }}}{{{{(2\pi )}^2}}}}{\mkern 1mu} {i^{ - l}}{e^{il{{\phi '}_k}}}\frac{{2\pi }}{\kappa }{\mkern 1mu} \delta ({{k'}_ \bot } - \kappa ){e^{i{{\vec k'}_ \bot } \cdot \vec x}}}  \times
    \\
    \sum\limits_{s = 0}^{v - 1} {\sum\limits_{t = 0}^{v - 1} {{a_{st}} \cdot \delta \left( {\vec x - {{\vec p}_{m'n'}} - {{\vec p}_{st}}/v - {{\vec k'}_ \bot }(j + u/v){z_T}/k} \right)} }
\end{multline}

It can be seen that in the fractional planes,  a group of “images”, the number of which is determined by the values of the numbers $u$ and $v$ for which the coefficients $a_{st}$ are not equal to zero, corresponds to each opening of the grating. Apparently,  the number of images is to be the same as in the case of the classical Talbot effect. Calculations of the coefficients can be found in articles \cite{Case2009,Berry,Arrizon}. In this respect, the results should be consistent with the classic Talbot effect. The essential difference for the images obtained by illuminating the arrays of the Bessel beam is the dependence of the radii of the rings on all three parameters $j$,  $u$, and $v$.

The expression of \eqref{5:17} shows that the "images" are a set of rings  with the radii
\begin{equation}\label{5:18}
   {\rho_{juv}} = \kappa \frac{{{p^2}}}{\pi }{\mkern 1mu} \left( {j + \frac{u}{v}} \right)
\end{equation}
and each multiplied by the factor ${{a_{st}}}$. The radius of the rings grows up with $j$ and $\beta$. It is proportional to the transverse wave number of the incident Bessel beam and to the square of the lattice period, and does not depend on the value of the topological charge of the incident beam $l$. The exponential factors in \eqref{5:17} describe the phase distribution in these rings, i. e. the annular beams formed in the Talbot planes retain the topological charge of the illuminating Bessel beam.

\subsection{Condition for existence of annular beamlets}

It is clear, that we can observe the rings only if they do not overlap each other. If the diameter of the rings exceeds the distance between their centers, the ring structure collapses. The formula of \eqref{5:18} gives the size of ideal infinitely thin ring. For real gratings with holes of a finite radius, the rings have a finite width, but, as calculations and experiments show, even if the ratio of the hole diameter to the period is equal to 2, the width of the rings remains relatively small. If we neglect the ring width, the condition of the existence of annular light beams can be written as
\begin{equation}\label{6:01}
   \kappa  \cdot \frac{{{p^2}}}{\pi } \cdot \left( {j + \frac{u}{v}} \right) \leqslant \frac{p}{{2 \cdot w}},
\end{equation}
where $w$ is the factor of multiplication of the ``images '' for a particular Talbot plane. For instance, $w = 1$ for the main Talbot planes ($u=0$) and the half-Talbot planes ($u/v=1/2$), whereas  $w = 2$ for the  planes with $u/v=1/4$ and $3/4$ and $w = 3$ for $u/v=1/3$ and $2/3$.

\begin{figure}[htbp]
\centering
\fbox{\includegraphics[width=7.5cm]{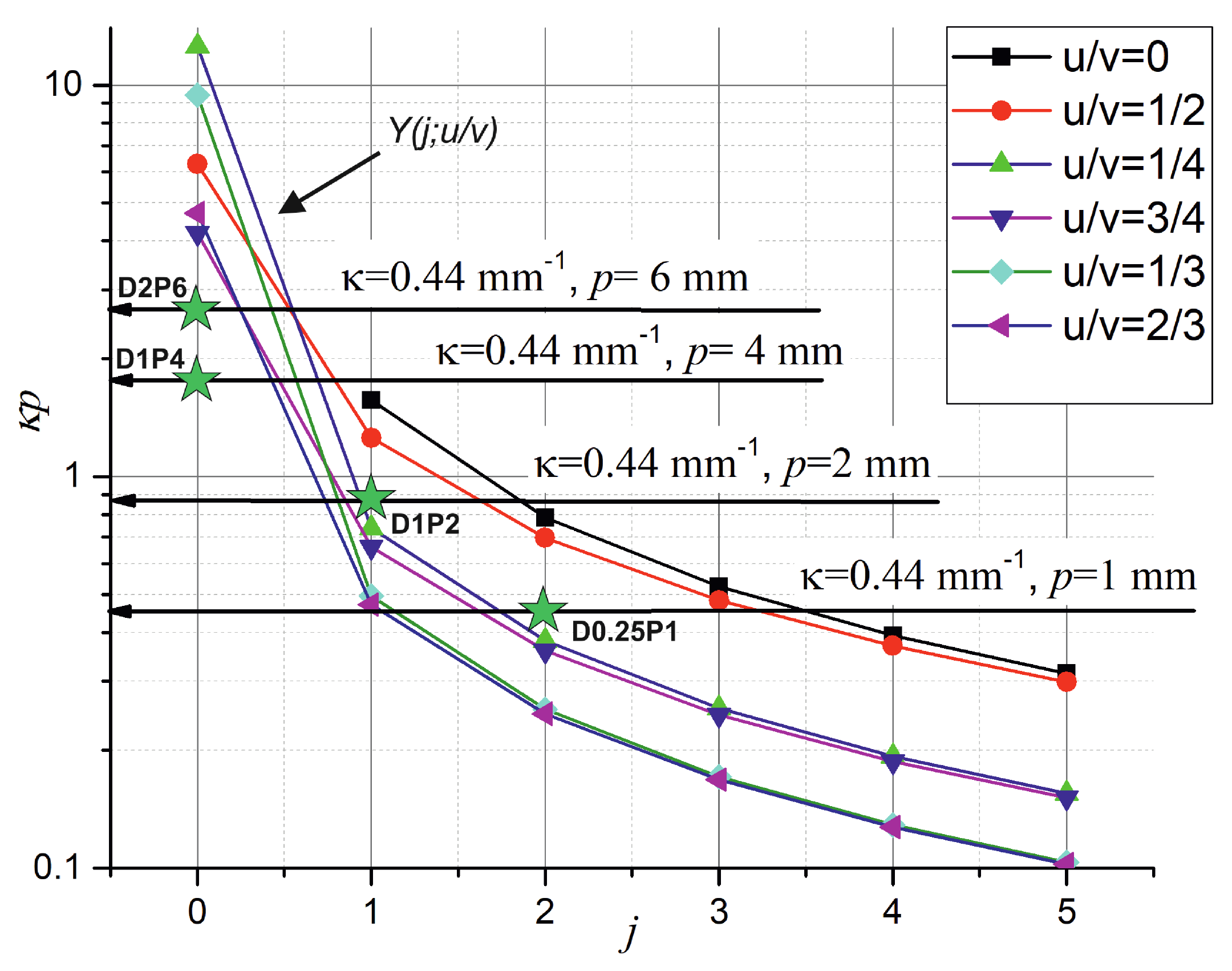}}
\caption{Conditions for forming periodic lattices of annular beams in Talbot planes. The lattices are formed if the product of the transverse wavenumber of the Bessel beam and the grating period, $\kappa p$,  lies below the curves $ Y(j;u,v)$ satisfying  \eqref{6:02}. The asterisks denote values of $\kappa p$ corresponding to the experimental results shown in Fig.~\ref{fig:comparison_1}.
}
\label{fig:Product_kappa_p}
\end{figure}

From \eqref{6:01}  it follows that the region of existence of the rings depends on the grating period and the diameter of the illuminating Bessel beam and can be written as
\begin{equation}\label{6:02}
   \kappa p \left( {j,u,v} \right) \leqslant  \frac{\pi }{{2 \cdot w(u,v) \left( {j + u/v} \right)}} = Y(j;u,v).
\end{equation}
\begin{figure}[htbp]
\centering
\fbox{\includegraphics[width=7.5cm]{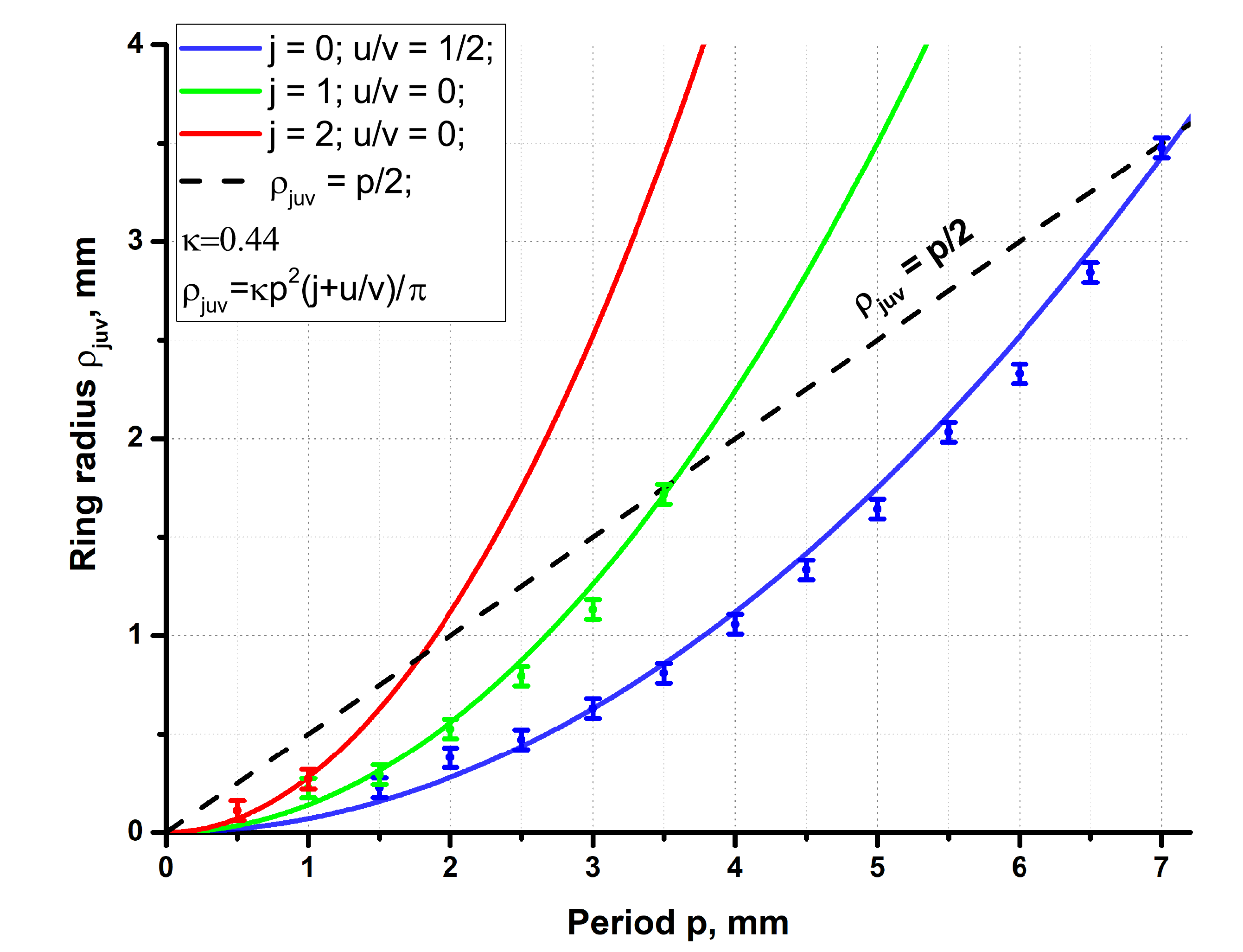}}
\caption{
Annular dependence of beam radii vs. grating period in the main and fractional Talbot planes calculated using \eqref{5:10}. The dashed line shows the maximum achievable radius for each period.
}
\label{fig:Condition_of_existence}
\end{figure}

The upper boundaries of the regions of existence of annular beams $ Y(j;u,v)$ are drawn in Fig.~\ref{fig:Product_kappa_p}. The arrows show the $\kappa p$ values corresponding to the parameters of the experiments (see. Section \ref{sec:Exp_and_Sim}). In Fig.~\ref{fig:Product_kappa_p}, only the points corresponding to integer values of $j$  have meaning, and the straight lines are drawn for the convenience only. If at a certain period of the lattice the rings of the diffraction pattern overlap, the radii of the rings can be reduced by expanding the incident beam with a telescopic system that decreases $\kappa$.

The conditions for the existence of ring structures for several Talbot planes are shown in Fig.~\ref{fig:Condition_of_existence}. It is seen that with the increase in the number of the main Talbot plane, it becomes more difficult to form a lattice of annular microbeams.

\section{Experiment and simulation}
\label{sec:Exp_and_Sim}

\begin{figure}[htbp]
\centering
\fbox{\includegraphics[width=\linewidth]{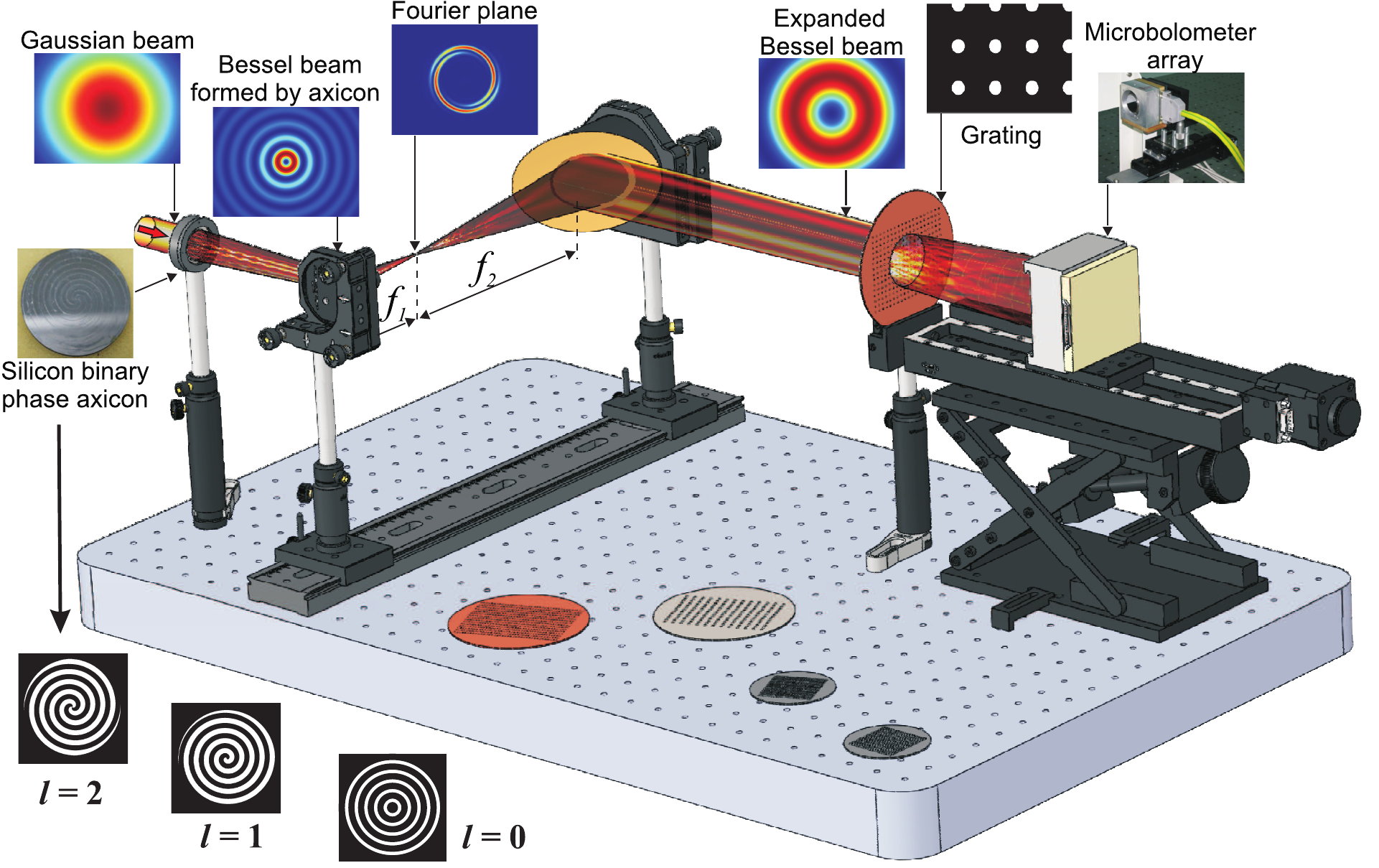}}
\caption{Experimental setup for study of the diffraction of Bessel beams on amplitude gratings. Spiral phase binary axicon transforms Gaussian beam to the Bessel beam, which is expanded by telescopic system and illuminates amplitude grating. The diffracted field is recorded by a $320\times 240$ microbolometer array.}
\label{fig:ExpSchematic}
\end{figure}

To verify the results of the analytical studies, we carried out experiments and numerical simulations. The experiments were performed at the wavelength $\lambda= 141$~$\mu$m using the setup shown in Fig.~\ref{fig:ExpSchematic}, with an optical system slightly modified as compared with \cite{Knyazev2018OE}. The Gaussian beam  of the Novosibirsk free electron laser was transformed into a Bessel beam of zero, first, or second order (the topological charges $l = 0,$~$1,$~$2$) using silicon binary axicons with circular or spiral zones \cite{Pavelyev}. The Bessel beam was expanded  4.5 times with a telescopic system consisting of two parabolic mirrors. The expanded beam illuminated an amplitude grating of round holes with a diameter $D$ and a period $P$. The radial wavenumber of all the beams $\kappa$ was equal to  $0.44$~mm$^{-1}$ (see \cite{PRA}). The diffraction pattern behind the grating was recorded by a microbolometer array \cite{Demyanenko}  moved along the optical axis by a motorized translation stage. The numerical simulation was performed within the framework of the scalar theory of diffraction using the WaveThruMasks program written in Matlab \cite{Kameshkov2019}.

\begin{figure}[htbp]
\centering
\fbox{\includegraphics[width=7.5cm]{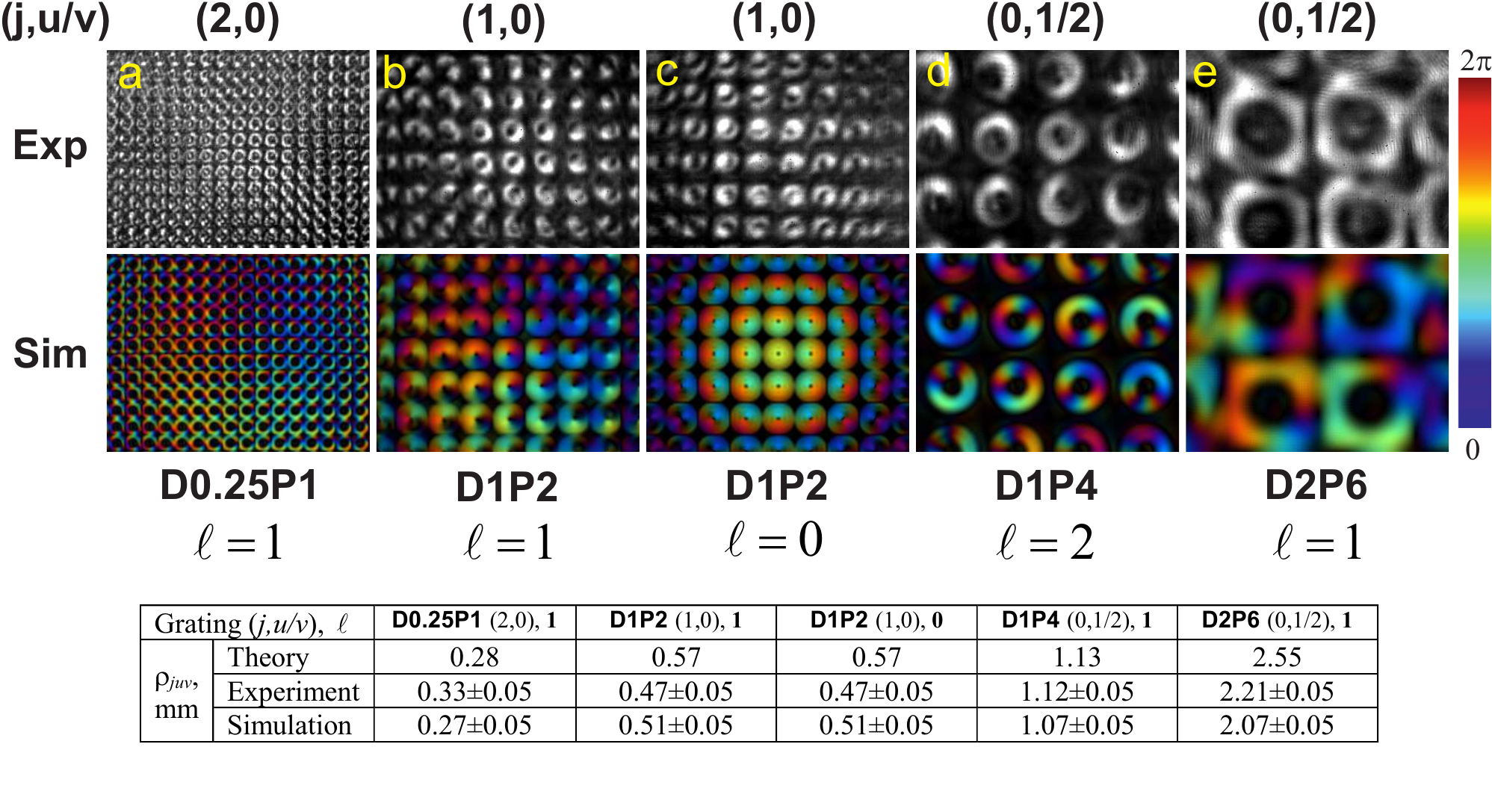}}
\caption{Examples of diffraction patterns obtained experimentally along with results of numerical calculations for Bessel beams of zero and first orders. Artificial colors show phase distribution. Results were obtained at wavelength $\lambda  = 0.141$ $\mu$m. Image size is $16.32 \times 12.24$ mm. }
\label{fig:comparison_1}
\end{figure}

 In the experiments, we used both conductive and nontranparent to radiation dielectric 2D gratings with periods of 1 to 6~mm and openings of 0.25~ to 2~mm. Selected diffraction patterns obtained  experimentally along with results of both numerical and analytical calculations are presented in Fig.~\ref{fig:comparison_1}. The results clearly shows that the ring radii in the main and fractional Talbot planes do not depend on the illuminating beam topological charge.  The values of radii calculated analytically and obtained experimentally and numerically well agree with each other.


It should be emphasized that the pattern shown in Fig.~\ref{fig:comparison_1}a was obtained for the grating, the diameter of the holes in which is equal to only two wavelengths of the incident radiation. Although formally speaking, under these conditions, the scalar theory of diffraction is not applicable, nevertheless, the results of numerical calculations are in excellent agreement with the experiment. Note also that the patterns were identical for both metal and dielectric gratings. In all frames of Fig.~\ref{fig:comparison_1}, the rings formed in the Talbot planes are clearly visible, because the condition of \eqref{6:01} is satisfied. The values of the parameter $\kappa p$ corresponding to these pictures are shown in Fig.~\ref{fig:Condition_of_existence} with asterisks. Due to the broadening of the rings, however, their shape can be distorted, even if the formal condition of existence is still fulfilled, as can be seen, for example, on the frame obtained with the D2P6 lattice of the ring.

\section{Discussion}

In the first sections of this paper we solved the problem of the Bessel beam diffraction  on a rectangular grating of small holes.  The electric field at an arbitrary point behind the grating is described by \eqref{5:04}. The integral in this expression cannot be taken in the general form, however, following the analogy with the diffraction of the plane wave on a periodic grating, it is reasonable to assume that in our case periodic structures with periods multiple to the grating period can also be formed in the Talbot planes.
\begin{figure}[htbp]
\centering
\fbox{\includegraphics[width=7.5cm]{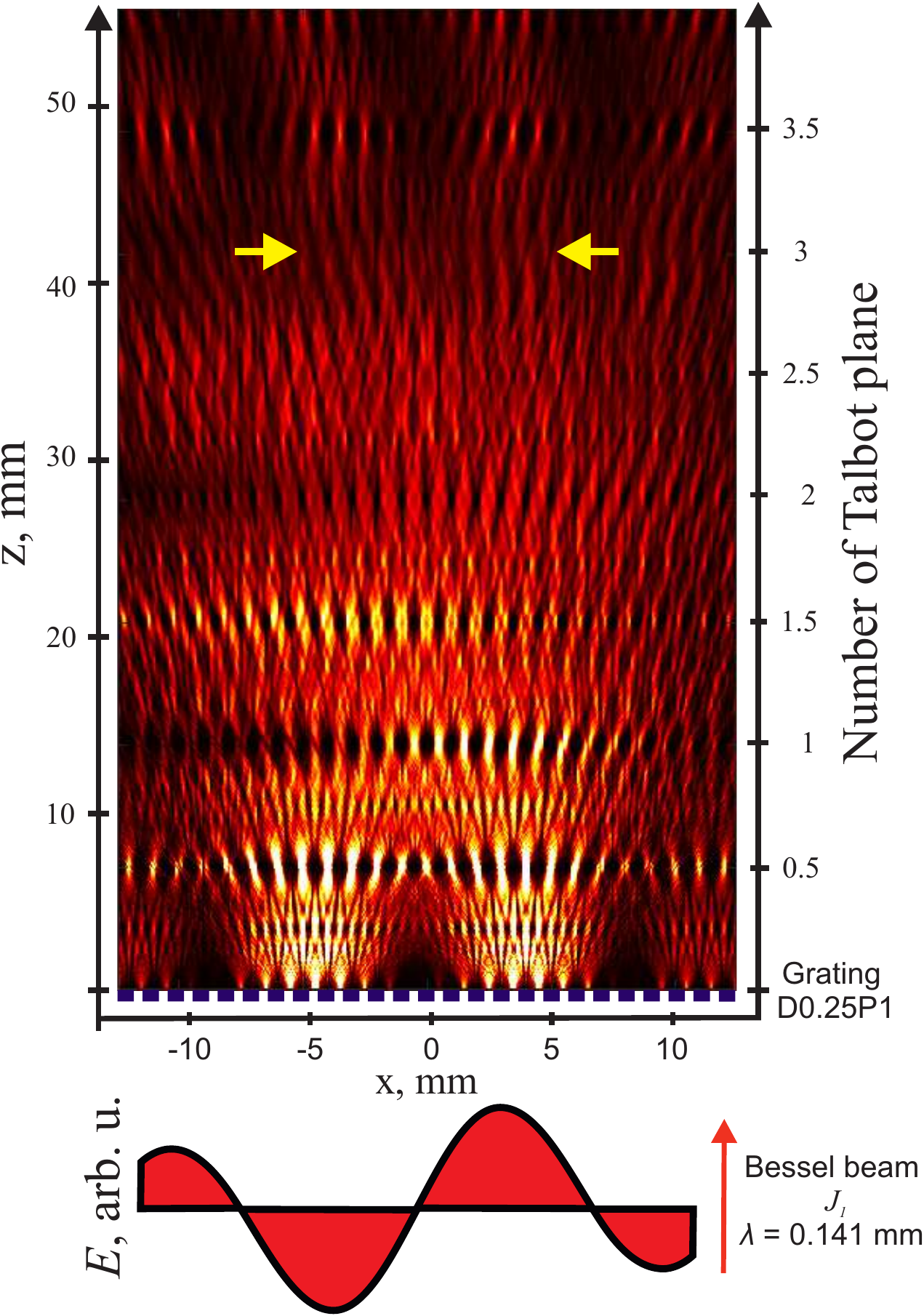}}
\caption{The Talbot carpet calculated for the Bessel beam of the first order ($l = +1$) diffracted on a rectangular grating of round openings 0.25~mm in diameter. Grating period is 1~mm. Transverse wavenumber of the Bessel beam is 0.44~mm$^{-1}$. The yellow arrows shows the borders of the frame presented in Fig.~\ref{fig:comparison_1}a.
}
\label{fig:carpet}
\end{figure}

Indeed, from the expressions for the distribution of the electric field in the main Talbot plane of \eqref{5:08}, the half Talbot plane of \eqref{5:13} and an arbitrary fractional Talbot plane of  \eqref{5:17}, it can be seen that a certain periodic structure of the diffracted wave is observed. Unlike the classical Talbot effect, we observe not the self-images of the lattice holes, but rings, the diameters of which depend on the grating period $p$,  the transverse wavenumber of the Bessel beam $\kappa$, and the indexes $j,u,v$ which define the Talbot plane.  We note here that the diffraction pattern in the form of single rings, whose diameter does not depend on the order of the Bessel beam $ l $, but depends only on the wavenumber $ \kappa $, indicates that these rings are essentially Fourier transforms of conically converging plane waves, into which Bessel beams can be decomposed.

However, the complexity of the diffraction pattern is determined not only by the dependence on a larger number of parameters but also by the fundamental difference in the geometry. In the case when the grating is illuminated by a wave whose phase is constant (for example, by a plane or a Gaussian wave), the self-image of the openings appears already in the nearest Talbot planes and repeats with increasing distance until all diffraction orders disperse due to the limited aperture. From this it follows that, even if the beam is axisymmetric, that affects the intensity distribution, but does not change fundamentally the diffraction pattern.

In the case of Bessel beams, the phase of the incident beam has, in general, both radial and azimuthal dependences, and for higher-order beams, the intensity near the axis is zero. Already the latter circumstance indicates that the diffraction pattern near the axis is finally formed only at a great distance, when the diffraction orders that come from opposite sides of the beam illuminating the grating begin to overlap. If we carefully consider the phase distribution in Fig.~\ref{fig:comparison_1}, obtained by numerical simulation of diffraction patterns, it is easy to see that in the rings located near the axis, the phase distribution corresponds to the azimuthal phase distribution in the illuminating beam ($\Phi = l\phi$). At a distance from the optical axis, the dominance of the phase of the illuminating beam as a whole becomes apparent, which can be explained by the influence of the zero diffraction order. In this case, the magnitude and sign of the topological charge of the microbeams are generally conserved, but the phase incursion in azimuth becomes not monotonous. In the analytical solutions, this phase shift, which depends on the transverse coordinate, is related to the phase factor ${\vec k'}_\bot  \cdot \vec x$ in the double integral under the sum sign.

The above reasoning is clearly illustrated in Fig.~\ref{fig:carpet}, in which the diffraction pattern shown in Fig.~\ref{fig:comparison_1}a is presented in the form of a Talbot carpet, that is, a cross-section of the diffraction field in the $xz$ plane. The classical Talbot effect ( self-images of the holes) dominates in the Talbot plain (0,1/2), in which the wavelets from the opposite part of the first Bessel ring does not interfere. In the planes (1,0) and (1,1/2), the diffraction pattern becomes asymmetric, and a regular pattern of annular microbeams is forming in the planes (2,1/2) and (3,0) near the optical axis. It is easy to calculate that the Huygens waves emerging from the opposite parts of the first Bessel ring intersect an the distance $z \sim \Delta x/(\lambda / d)=5/0.14 \approx 40$~mm, which is consistent with Fig.~\ref{fig:carpet}.

\section{Conclusion}
In this paper, we derived analytical expressions describing the diffraction of Bessel beams on a periodic two-dimensional lattice, and confirmed the results obtained by numerical calculations and experiments. Although the diffraction pattern does not reproduce grating images, the fact that the periodic system of rings is observed at the classical Talbot distances makes it possible to classify this work as a new branch in the theory of the Talbot effect. By optimizing the parameters of the gratings and Bessel beams, one can obtain in the paraxial region a set of ring beams with a given topological charge, which can be used to create an optical trap grating or transfer the mechanical moment to a sequence of microturbines, for example, in microfluidic systems. Another possible application is the excitation of vortex surface plasmon polaritons by the fire-wire technique on a Lattice of cylindrical conductors in plasmonic devices.

\section*{Funding Information}
Russian Science Foundation (RSF) (19-12-00103).

\begin{acknowledgments}
The experiments were carried out at the Novosibirsk free electron laser (NovoFEL) facility using the  equipment of the Siberian Synchrotron and Terahertz Radiation Center. The authors thank the NovoFEL team for technical support, B.~G.~Goldenberg for production of the gratings with small openings, V.~S.~Pavelyev for the design and fabrication of the binary axicons, and Yu.~Yu.~Choporova, G.~N.~Kulipanov, N.~A.~Vinokurov for the helpful discussions.
\end{acknowledgments}

\appendix

\section{Appendixes}\label{app:Poisson}

Recall the Poisson summation rule (see, for example, \cite{Sidorov1989}). Suppose we need to calculate the sum of a row
\begin{equation}
 \label{8:01}
 \sum\limits_{m =  - \infty }^\infty  f (m).
\end{equation}
Define the function
\begin{equation}
\phi (\mu ) = \sum\limits_{m =  - \infty }^\infty  f (m + \mu ).
\end{equation}
It is periodic with a period of $ 1 $, so it can be expanded into a Fourier series:
\begin{equation}
\phi (\mu ) = \sum\limits_{m' =  - \infty }^\infty  {{\phi _{m'}}} {e^{2\pi im'\mu }},
\end{equation}
where
\begin{multline}
    \label{8:04}
 {\phi _{m'}} = \int_0^1 {\text{d}} \mu {e^{ - 2\pi im'\mu }}\phi (\mu ) =
 \\
 = \int_0^1 d \mu {e^{ - 2\pi im'\mu }}\sum\limits_{m =  - \infty }^\infty  f (m + \mu ) =
 \\
 = \sum\limits_{m =  - \infty }^\infty  {\int_m^{m + 1} d } \mu {e^{ - 2\pi im'\left( {\mu  - m} \right)}}f(\mu ) =
 \\
    = \sum\limits_{m =  - \infty }^\infty  {\int_m^{m + 1} d } \mu {e^{ - 2\pi im'\mu }}f(\mu ) =
 \\
 = \int_{ - \infty }^\infty  {\text{d}} \mu {e^{ - 2\pi im'\mu }}f(\mu ).
\end{multline}
Note that
\begin{equation*}
\sum\limits_{m =  - \infty }^\infty  f (m) = \phi (0) = \sum\limits_{m' =  - \infty }^\infty  {{\phi _{m'}}}.
\end{equation*}
Substituting the expression for $\phi_{m'}$ here, we get
\begin{equation*}
 \sum\limits_{m =  - \infty }^\infty  f (m) = \sum\limits_{m' =  - \infty }^\infty  {\int_{ - \infty }^\infty  d } \mu {e^{ - 2\pi im'\mu }}f(\mu ).
\end{equation*}
Replacing $\mu$ with $m$, we get
\begin{equation}
\sum\limits_{m =  - \infty }^\infty  f (m) = \sum\limits_{m' =  - \infty }^\infty  {\mkern 1mu}  \int\limits_{ - \infty }^\infty  d m{e^{ - 2\pi im'm}}f(m).
\end{equation}
The generalization to the two-dimensional case is obvious:
\begin{equation}
\label{8:08} 
 E(x,z) = \sum\limits_{m'n'} {\int\limits_{ - \infty }^\infty  {\int\limits_{ - \infty }^\infty  {dm} } \int\limits_{ - \infty }^\infty  {\int\limits_{ - \infty }^\infty  {dn} } } {\mkern 1mu} {E_{mn}}{e^{ - 2\pi im'm - 2\pi in'n}},
\end{equation}


\end{document}